\documentstyle[preprint,pra,aps]{revtex}
\tightenlines
\begin{document}
\title{ 
Nonlinear Enhancement of the Multiphonon Coulomb Excitation 
in Relativistic Heavy Ion Collisions
}
\author{ M.S. Hussein, A.F.R. de Toledo Piza, O.K. Vorov}
\address{ 
Instituto de Fisica, 
Universidade de Sao Paulo \\
Caixa Postal 66318,  05315-970,  \\
Sao Paulo, SP, Brasil 
}
\date{13 August 1998}
\maketitle
\begin{abstract}
We propose a soluble model to incorporate the nonlinear effects in the
transition probabilities of the multiphonon Giant Dipole Resonances
based on the SU(1,1) algebra. Analytical expressions for the
multi-phonon transition probabilities are derived. Enhancement of the
Double Giant Resonance excitation probabilities 
in relativistic ion collisions scales as 
$(2 k +1)(2k)^{-1}$ for the degree of nonlinearity $(2k)^{-1}$                
and is able to reach values $1.5-2$ compatible with experimental data.
The enhancement factor is found to decrease with increasing 
bombarding energy.

[KEYWORDS: Relativistic Heavy Ion Collisions,Double Giant Resonance] 
\end{abstract}

\newpage
Coulomb Excitation in collisions of relativistic ions is one of
the most promising methods in modern nuclear physics 
\cite{EMLING,BCHTP,bbabs,CH-FR,HBCH}. 
One of the most interesting applications of 
this method to studies of nuclear structure is the possibility to 
observe and study the multi phonon Giant Resonances \cite{EMLING}.
In particular, the double Dipole Giant Resonances (DGDR) have been 
observed in a number of nuclei 
\cite{SCH-EXP-Xe,R-EXP-Pb,AU-EXP-Au}. 
The ``bulk properties''
of the one- and two-phonon GDR are now partly understood \cite{EMLING}
and they are in a reasonable agreement with the theoretical picture
based on the concept of GDR-phonons as almost harmonic quantized vibrations.

Despite that, there is a persisting discrepancy between the theory and the
data, observed in various experiments 
\cite{SCH-EXP-Xe,R-EXP-Pb,AU-EXP-Au,B-208-Pb-1.33,B-208-Pb-LOW} that still 
remains to be understood: the double GDR excitation cross sections 
are found enhanced by factor $1.3-2$ with respect to the predictions
of the harmonic phonon picture 
\cite{EMLING},\cite{bbabs},\cite{AW65},\cite{WA79}.
This discrepancy,
which almost disappears at high bombarding energy,
has  attracted much attention in current literature 
\cite{CH-FR},
\cite{CRHP94,Norbury-Baur,hot-phonon1,hot-phonon2,BZ93},
\cite{vccal,vca2,B-D};
among the approaches to resolve the problem are the higher order
perturbation theory treatment \cite{BZ93},  and studies of 
anharmonic/nonlinear aspects of GDR dynamics \cite{CH-FR},
\cite{vccal,vca2},\cite{B-D}.
Recently, the concept of hot phonons \cite{hot-phonon1},\cite{hot-phonon2}
within Brink-Axel mechanism was proposed that provides microscopic 
explanation of the effect. 
These seemingly orthogonal explanations deserve clarification which 
we try to supply here.

The purpose of this work is to examine, within a a soluble model 
the role of the
nonlinear effects 
on the transition amplitudes
that connect the multiphonon states in a heavy-ion Coulomb excitation 
process. Most studies of anharmonic corrections \cite{vccal,vca2,B-D}
concentrated on their 
effect in the spectrum 
\cite{CCG,Bertsch-Feldmeier}.
Within our model, the nonlinear effects are described by a single 
parameter, and the model contains the harmonic model as its limiting case
when the nonlinearity goes to zero. 
We obtain analytical expressions for the probabilities of excitation
of multiphonon states which substitutes the Poisson formula of the 
harmonic phonon theory. 
For the reasonable values of the nonlinearity, the present model
is able to describe the observed enhancement of the double GDR
cross sections quoted above.
 
Having in mind to show how analytical results follows from the
nonlinear model, and to explain how the model works, we restrict 
ourselves
here to its simplest version (transverse approximation, or SU(1,1)
dynamics) and keep numerics up to minimum level.
We postpone till further publications detailed numerical analysis
and comparison with the data. Microscopic origins of the nonlinear 
effects (which are considered here phenomenologically) are also
beyond the scope of this presentation. 

We work in a semiclassical approach \cite{AW65} to the
coupled-channels problem, i.e.,  
the projectile-target relative
motion is approximated by a classical trajectory and the
excitation of the Giant Resonances is treated 
quantum mechanically \cite{CRHP94}.
The use of this method is justified due to the small wavelenghts
associated with the relative motion 
in relativistic heavy ion collisions. 
The separation coordinate is treated as a classical time 
dependent variable, and the projectile motion 
is assumed to be a straight line \cite{BCHTP}.

The intrinsic
dynamics of excited nucleus is governed by a time dependent quantum 
Hamiltonian
(see Refs. \cite{AW65},\cite{WA79}).
The intrinsic state $\vert \psi(t)>$ of excited nucleus 
is the solution of the time dependent Schr\"odinger equation
\begin{equation} \label{SCHR}
i {\partial \vert \psi(t)> \over \partial t} =
\left[ H_0 + V (t) ) \right]\ \vert \psi(t)> 
\end{equation}
where  $H_0$ is the intrinsic Hamiltonian and $V$ is the
channel-coupling interaction. We use the system of units where 
$\hbar=1$, $c=1$.
The standard 
coupled-channel problem for the amplitudes $a_n(t)$ reads
\begin{equation} \label{expan}
\vert \psi (t)> = \sum_{n=0}\ a_n(t)\ \vert n>\ \exp \Big(-i
E_n t \Big),
\end{equation}
where $E_n$ is the energy of the state $\vert n>$ in the wave packet
$|\psi \rangle$. In our treatment, the nuclear states are specified by the
numbers of excited GDR phonons, $N$ or $n$.
Taking scalar product
with the states $<N \vert $, we get the set of coupled equations
for the amplitudes $a_n$ as functions of impact parameter $b$
\begin{equation} \label{CC}
i  {\dot a}_N(t) = \sum_{n=0} < N\vert V\vert n >
e^{i(E_N-E_n) t } a_n(t).
\end{equation}
We assume the colliding nuclei to be in their ground states 
before the collision. 
The amplitudes obey the initial condition $a_n(t\rightarrow -\infty)
= \delta(n,0)$ and they tend to constant values as
$t\rightarrow \pm \infty$ (the interaction $V(t)$ dies out at 
$t\rightarrow \pm \infty$). 
The excitation probability of an intrinsic
state $\vert N> $ in a collision with impact parameter $b$ is given as
\begin{equation} \label{Pn}
W_N(b) = \vert a_N(\infty)\vert ^2 .
\end{equation}
and the 
total cross section for excitation of the state $\vert N>$ is given by the
integral over the imact parameter
\begin{equation} \label{sigman} 
\sigma_N= 2 \pi \int\limits_{b_{gr}}^{\infty} b W_N(b)  db  
\end{equation}
with the grazing value $b_{gr}= 1.2 (A_{exc}^{1/3}+A_{sp}^{1/3})$ as the
lower limit. Hereafter, the labels $exc$ ($sp$) refer 
to the excited (spectator) partner in a colliding projectile-target pair. 
We neglect the here nuclear contribution 
\cite{HRBHPT} to the excitation process.

In the following, it is convenient to treat
the coupled channel equations (\ref{SCHR}),(\ref{CC})
in terms of the unitary operator $U_I$ (in the interaction representation)
that acts in the reference 
basis of multiphonon states (including the ground state $|0\rangle$):
\begin{equation} \label{U}
i \frac{d} {d t} U_I(t) = V_I(t)  U_I(t) , 
\qquad V_I(t)= e^{iH_0 t} V(t) e^{-iH_0 t}, 
\qquad U_I(t=-\infty) = I,
\end{equation}
and the time-dependent Hamiltonian $H(t)=H_0+V(t)$ that acts in the intrinsic
multi-GDR states is given by 
\begin{eqnarray} \label{BASIC}
H_0 = \omega N_d, \qquad N_d \equiv \sum\limits_{m}d^+_{m} d_{m},
\qquad \qquad \qquad 
\nonumber\\
V(t) = v_1(t) [ (E1_{-1})^{\dagger} - (E1_{+1})^{\dagger} ] +
v_0(t)   (E1_{0})^{\dagger} + Herm.Conj.  
\end{eqnarray}
where $E1_{m}^{\dagger}$ and $E1_{m}$ 
are the dimensionless operators acting in the
internal space of the multi-GDR states. In the harmonic 
approximation, they are given by the GDR phonon creation 
and destruction operators of corresponding angular momentum projection $m$, 
$E1_{m}^{\dagger}=d^{\dagger}_m$.
The function $v_1(t)$ is given \cite{BCHTP} by
\begin{eqnarray} \label{v1}
v_1(t) = \frac{w}{[1 + (\frac{\gamma v}{b}t)^2]^{3/2}}, 
\quad w=   \rho \frac{Z_{sp} e^2 \gamma}{2 b^2} 
\sqrt{ \frac{ N_{exc} Z_{exc} }{A_{exc}^{2/3} m_N \cdot 80 MeV} } ,
\end{eqnarray}
(the corresponding expression for $v_0(t)$ can be found in Ref.
\cite{BCHTP}). Here, $m_N$ and $e$ are the proton mass and charge,
$Z$, $N$ and $A$ denote the nuclear charge,
the neutron number and the mass number of the colliding partners,
$\gamma=(1-v^2)^{-1/2}$ is relativistic factor, $v$ is the velocity and
the parameter $\rho$ is the deal of the strength absorbed by the
collective motion (usually assumed to be close to unity) \cite{EMLING}.

The harmonic approximation (ideal bosons) yields the transition 
probabilities between the states with numbers of GDR phonons differing
by unity to grow linearly $\propto N$.
This model of ideal bosons has well known exact solution
(see, e.g. \cite{WA79}) 
that is given by the Poisson formula for the excitation probabilities 
\begin{eqnarray} \label{POISSON}
W_N = e^{-|\alpha^{harm}|^2} \frac{|\alpha^{harm}|^{2N} }{N!},
\nonumber\\
|\alpha^{harm}|^2 
= \sum_{m=0,\pm 1} |\alpha^{harm}_m|^2 = 
2 |\alpha^{harm}_1|^2 + |\alpha^{harm}_0|^2
\end{eqnarray}
where the amplitudes $\alpha^{harm}_m$ are expressed in terms of the 
modified Bessel functions. At the colliding energies sufficiently high,
the longitudinal contribution ($\propto |\alpha^{harm}_0|^2$) is
suppressed by a factor proportional to $\gamma^{-2}$ 
\cite{bbabs}.
In the following, we will work in the ``transverse approximation''
dropping the longitudinal term for the sake of simplicity. This is a
good approximation at high energies and the results are still 
qualitatively valid at lower energies. 

Our idea is to keep the spectrum of GDR system harmonic with the
Hamiltonian $H_0 = \omega N$. That is supported by the systematics
of the observed DGDR energies, $E_2$, which yields 
$E_2 \simeq (1.75-2) \omega$ \cite{EMLING}.
The conclusion on the weak anharmonicity in the spectrum follows also
from theoretical considerations \cite{CCG},\cite{Bertsch-Feldmeier}.

The transition operators $E1^{\dagger}, E1$ can however include 
nonlinear effects. (In particular, this could be a result of 
nonlinearities in the phonon Hamiltonian obtained 
in higher orders of perturbation theory).
A reasonable accounting for these nonlinear effects that 
we adopt in this work consists of generalization of the transition
operators
\begin{equation} \label{E1}
E1_{m}^{\dagger}=d^{\dagger}_m \sqrt{1 + \frac{1}{2k} N_d}, \quad 
E1_{m}=\sqrt{1 + \frac{1}{2k} N_d} \quad d^{\dagger}_m , \qquad
\end{equation}
where the parameter $k \neq 0$ determines the strength of 
the nonlinear effects,
i.e., the problem reduces to the harmonic oscillator 
with linear coupling when 
\begin{equation}
\frac{1}{ 2 k} \rightarrow 0.
\end{equation}
If $k<0$, the transition probabilities are suppressed as $N$ grows.
Positive values of $k$ that we will consider here correspond
to the enhancement of the matrix elements.

It is convenient to introduce the three following operators
$D^+$,  $D^-$ and $D^0$
\begin{eqnarray} \label{SU11}
D^+ = \frac{1}{2^{1/2}} (d^+_{+1} - d^+_{-1}  )\sqrt{ 2k + N_d}, \quad
D^- = \frac{1}{2^{1/2}} \sqrt{ 2k + N_d} (d_{+1} - d_{-1}  ), \quad
\nonumber\\ 
D^0 = \frac{1}{4} 
\left[ (d^+_{+1} -  d^+_{-1} )(d_{+1} - d_{-1} ) + 2(2k+N_d) \right], \qquad
\end{eqnarray}
with $ N_d \equiv  d^+_{+1} d_{+1} + d^+_{-1} d_{-1}$. It is easy to check
that they
obey the standard commutation relation for the {\it noncompact} 
SU(1,1) algebra
\begin{equation} \label{ALGEBRA}
\left[ D^- , D^0 \right] = D^-, \quad \left[ D^+ , D^0 \right] = -D^+ , 
\quad \left[ D^- , D^+ \right] = 2 D^0.
\end{equation}
The dynamics of the our system, in transverse approximation,
can be written in terms of the operators $D^+$,  $D^-$ and $D^0$ 
(\ref{SU11}) only.
In the interaction representation, the evolution equation 
(\ref{U}) takes the form
\begin{eqnarray} \label{INTERACTION}
i \frac{d} {d t}U_I(t) = \left[ 
\frac{v_1(t)}{\sqrt{k}} e^{i \omega t}  D^{\dagger} + 
\frac{ v_1(t)}{\sqrt{k}} e^{- i \omega t} D^- 
\right] U_I(t),   
\end{eqnarray}
for $U_I(t)$
(the last equation follows from 
(\ref{U}), (\ref{BASIC}) after using the commutation relations 
(\ref{ALGEBRA}) and $[ N_d, D^{\pm} ] = \pm D^{\pm}$.

From purely mathematical viewpoint,
the problem described by the last equation
drops into the universality class of the systems
with SU(1,1) dynamics that can be analyzed by means of generalized 
coherent states \cite{PERELOMOV}for the SU(1,1) algebra. 
In particular, the problem
of the parametric excitation of a one-dimensional 
harmonic oscillator (\cite{PERELOMOV},\cite{PP})
belongs to the same class. The physical meaning of the algebra 
generators (\ref{SU11}) in our case is of course quite different from that 
of \cite{PERELOMOV,PP}.
(For other algeabraic approaches to scattering problems,  
see Refs.\cite{GWI}, \cite{AGI}).

The formal solution of our problem is given by the expression for the unitary
operator $U_I(t)$ as a time-ordered exponential (see, e.g., \cite{KIRZHNITS})
\begin{equation} \label{T-EXPONENT}
U_I(t) = T 
exp \left( -i \int\limits_{-\infty}^{t} d t' V_I(t') \right) 
\end{equation}
Due to closure of the commutation relations between the operators 
$D^{+},D^{-}$ and $D^0$ that enter the exponential in Eq.(\ref{T-EXPONENT}),
the time-ordered exponential can be represented in another equivalent form
that involve ordinary operator exponentials only 
(see, e.g., \cite{KLEINERT}).
The operator $U_I(t)$ can be expressed as
\begin{equation}  \label{DECOMPOSITION}
U_I(t) = 
exp \left[ \frac{\alpha(t)}{\sqrt{k}} D^+ \right] 
exp \left[ \left[ log \left(1 - \frac{|\alpha(t)|^2}{k}\right) -i\phi(t) 
\right] D^0 \right]
exp \left[- \frac{\alpha^{*}(t)}{\sqrt{k}} D^- \right] 
\end{equation}
where one has the product of ordinary exponentials 
\cite{PERELOMOV} which contain
some time-dependent complex number $\alpha(t)$ 
(star means complex conjugation) and real number $\phi(t)$ (phase).
The functions $\alpha(t)$ and $\phi(t)$ can be found from 
simple differential equations 
which relate the unknown $\alpha(t)$ and $\phi(t)$ with the 
function $v_1(t)$ in the Hamiltonian $H(t)$. These equations (see below)
can be 
restored after substituting the right hand side of Eq.(\ref{DECOMPOSITION})
into the left hand side of the Schr\"odinger equation for the operator
$U_I(t)$ (\ref{INTERACTION}) and collecting the terms which have the same
operator structure. 

Proceeding this way, we obtain, after some algebraic manipulations, 
from (\ref{INTERACTION}) with
using the commutation relations (\ref{ALGEBRA}), the formula
\begin{displaymath}
\frac{d} {d t} e^{A} = \int\limits_0^1 
d \tau e^{ \tau A } \left( \frac{d} {d t} A \right)
e^{(1- \tau) A}
\end{displaymath}
for differentiating the operator exponential and the Baker-Hausdorf 
relations for the operator exponentials \cite{KIRZHNITS}, 
the following Riccati-type equation for the complex amplitude $\alpha$:
\begin{equation}   \label{RICCATI}
i \frac{d} {d t} \alpha = v_1(t) e^{i \omega t} +
v_1(t) e^{-i \omega t} \frac{\alpha^2}{k}.
\end{equation}
The expression for the phase $\phi(t)$ is given by a simple integral  
$\phi(t)= (2/k) 
\int\limits_{-\infty}^t dt_1 Re( v_1(t_1) \alpha(t_1) e^{-i \omega t_1})$ 
(as the phase does not contribute to the 
squared absolute values of amplitudes that enter the excitation 
probabilities and the cross sections, we will not be interested
in it in the following).

Once the solution for the differential equation (\ref{RICCATI}) is found, 
the expression for the amplitudes $a_N(t)$ which we are interesting in 
follows from (\ref{DECOMPOSITION}) immediately after projection of the
state 
\begin{displaymath}
|\psi(t)\rangle = U_I(t) |0 \rangle
\end{displaymath}
onto the intrinsic states with definite number
of GDR phonons, $N$. 
For example, we have From Eq.(\ref{DECOMPOSITION}) while acting onto the 
ground state a particularly simple result 
\begin{displaymath}
U_I(t) | 0 \rangle = 
e^{-i k \phi(t)} \left( 1 - \frac{|\alpha(t)|^2}{ k}  \right)^{k}
exp \left[ \frac{\alpha(t)}{\sqrt{k}} D^+ \right] | 0 \rangle. 
\end{displaymath}
We thus obtain the expression for the amplitude and the probabilities of the
transitions from the ground state to the excited states with
$N$ phonons:

\begin{eqnarray} \label{RESULT}
| a_{N}(\infty) | = 
\left( 1 - \frac{|\alpha( \infty)|^2}{ k}  \right)^{k} 
\left( 
\frac{ \Gamma(2k+N) }{ N! \Gamma(2k)} 
\right)
^{1/2} 
\left( \frac{|\alpha( \infty)|^2}{k} \right)^{N/2}
\nonumber\\
\quad  W_{N} = | a_{N}(\infty) |^2 \qquad \qquad \qquad \qquad \qquad
\qquad.
\end{eqnarray}
Here, the quantity $\alpha(+ \infty)$ is the asymptotic solution to
the Riccati equation (\ref{RICCATI}) at $t \rightarrow \infty$
subject to the initial condition $\alpha(- \infty) = 0$.
The simple nonlinear equation (\ref{RICCATI}) abbreviates all orders of 
quantum perturbation theory for the
problem Eqs.(\ref{U}),(\ref{BASIC}),(\ref{INTERACTION}). 
It is also seen from Eq.(\ref{DECOMPOSITION})
that unitarity is automatically preserved within present 
formalism ($U_I^{\dagger}=U_I^{-1}$).

The harmonic limit of these results corresponds to the case
$k \rightarrow \infty$, 
when the nonlinearity disappears in the transition operators (\ref{E1})
and the coupling to electromagnetic field becomes linear 
(harmonic approximation).
Then at $k \rightarrow \infty$ 
the last nonlinear term  drops from the  equation (\ref{RICCATI}),
and  the solution to the equation 
$|\alpha(\infty)|$ reduces to its harmonic value $|\alpha^{harm}_{\pm1}|$
given by modified Bessel functions \cite{WA79},\cite{EMLING}, 
At the same time, the expression for the probabilities $W$, 
Eq.(\ref{RESULT}),
reduces at $k \rightarrow \infty$ to the 
Poisson formula (\ref{POISSON}), 
thus the harmonic results \cite{WA79},\cite{EMLING} are restored.

The simple nonlinear equation (\ref{RICCATI})
for the amplitude $\alpha(t)$ can be 
readily solved numerically. One can construct the solution by means of
successive approximations, provided that the nonlinearity is small
enough to keep the first terms in the expansion to powers of
$|v_1|/\sqrt{k}$,
\begin{equation} \label{PERT-THEORY}
\alpha(t)=
-i \int\limits_{-\infty}^{t}
v_1(t_1)  e^{i \omega t_1} dt_1  +  
i \int\limits_{-\infty}^{t}
\frac{v_1(t_1)}{k}  e^{-i \omega t_1} dt_1 \left[ \int\limits_{-\infty}^{t_1}
v_1(t_2)  e^{i \omega t_2} dt_2 \right]^2 + ...
\end{equation}
In fact, the first term of this expansion is an excellent
approximation in many cases, especially for light projectiles/targets
when $|v_1|$ (\ref{v1}) is small.
We will use this approximation below as a good reference point to 
analyze the enhancement factor for excitation of DGDR. 
Keeping thus the first term in (\ref{PERT-THEORY}) we have 
\begin{equation} \label{FIRST-ORDER}
|\alpha(\infty)| \simeq
\left\vert -i \int\limits_{-\infty}^{\infty}
v_1(t)  e^{i \omega t} dt \right\vert  = 2 \frac{w}{\omega} \xi^2 K_1(\xi)
\end{equation}
i.e., the well known expression expression in terms of the 
modified Bessel function. While the amplitude $\alpha$ is thereby
reduced to the harmonic value in this approximation, 
the excitation probabilities  $W_{N}$ are still given by the formula for 
the nonlinear model, Eq.  (\ref{RESULT}).
Using (\ref{RESULT}) and (\ref{FIRST-ORDER}),
one has therefore the approximate expression 
for the excitation probabilities  
\begin{eqnarray}  \label{PERT-THEORY-W}
W_{N} = 
\left( 1 - \frac{ | K_1(\xi )|^2 s^2 }{ k}  \right)^{2k} 
\frac{ \Gamma(2k+N) }{ N! \Gamma(2k)} 
\left( \frac{ 2 | K_1(\xi) |^2 s^2 }{2k} \right)^{N},
\end{eqnarray}
here, $s = 2 \frac{w}{\omega} \xi^2$ and 
$\xi \equiv \frac{\omega b}{v \gamma}$.
The cross sections for the excitation of $N$-phonon GDR are
\begin{eqnarray}  \label{PERT-THEORY-S}
\sigma_{N} = 2 \pi \frac{ \Gamma(2k+N) }{ N! \Gamma(2k) (2k)^N} 
\int\limits_{b_{gr}}^{\infty} d b \left( 1 - \frac{ | K_1(\xi )|^2 s^2 }{ k}  
\right)^{2k} 
(2 | K_1(\xi ) |^2 s^2 )^{N}
\end{eqnarray}
At nonzero nonlinearity $1/(2k)>0$, the excitation 
probabilities $W_{N}$ 
for Multiple GDR turn out to be enhanced as compared to their 
values $W^{harm}_{N}$ in the harmonic limit. 
In general, deviation in the excitation probabilities 
for the $N$-phonon states $W_N$ (\ref{RESULT})
from their harmonic values $W^{harm}_{N}$ (\ref{POISSON})
are given by the ratio  
\begin{displaymath}
\frac{W_N}{ W^{harm}_{N} } = 
\prod\limits_{i=0}^{N-1}\left(\frac{2k + i }{2k}\right) 
\frac{( 1 - \frac{|\alpha(+ \infty)|^2}{k})^{2k}}
{e^{-2 |\alpha^{harm}_{1}|^2}} 
\frac{ | \alpha(+\infty) |^{2N}}{ | \alpha^{harm}_{1}(+\infty) |^{2N} }.
\end{displaymath}
In perturbative regime, when the amplitudes $\alpha$ (\ref{PERT-THEORY})
are close
to the harmonic solutions, the two last factors in this equation
are close to unity,
and the deviations are basically due to the factor 
$\prod\limits_{i=0}^{N-1}\frac{2k + i }{2k}$.
It is convenient to introduce the following ratio $R_2$ that characterizes
enhancement of the excitation probability of the two-phonon GDR
due to nonlinear effects with respect to the transition probability 
in the harmonic limit
\begin{equation} \label{R2}
R_2 \quad =  
\quad \frac{ W_{2} \quad / \quad W_{1}}
{ W^{harm}_{2} \quad / \quad W^{harm}_{1}}
\end{equation}
From equations (\ref{POISSON}),(\ref{RESULT}) we obtain the result 
\begin{equation} \label{R2a}
R_2 \quad =  \quad \frac{2k + 1 }{2k}
\frac{ | \alpha(+\infty) |^2}{ | \alpha^{harm}_{1}(+\infty) |^2}.
\end{equation}
The first factor in this expression, $\frac{2k + 1 }{2k}$, results
from the kinematic enhancement of the transition probabilities
due to nonlinear effects
considered here. 
When the amplitude $|\alpha(+ \infty)|$ is small, the second and the 
third factors in Eq.(\ref{R2a}) are close to unity and 
we obtain the result
\begin{equation}
R_2 \quad \simeq  \quad \frac{2k + 1 }{2k}.
\end{equation}

The cross sections can be obtained from the usual formula (\ref{sigman}).
For the ratio of the cross sections of excitation of the double GDR
calculated within present nonlinear model $\sigma_2$ and within
the harmonic phonon model $\sigma^{harm}_2$,  
we obtain in the same scaling regime
\begin{equation} \label{PSI}
r_2 \quad = \frac{\sigma_2}{\sigma^{harm}_2} 
\simeq
\quad \psi \frac{2k + 1 }{2k}
\end{equation}
where the exact numerical factor $\psi$ that stems from 
equations (\ref{RESULT}), (\ref{PERT-THEORY-W}), (\ref{PERT-THEORY-S}),
is close to unity when $|\alpha|\simeq |\alpha^{harm}_1| \ll 1$.
It contains dependence on the bombarding energy 
and describes deviations from the 
scaling behavior which are valuable at both low and high energies
(we discuss these deviations below).

The interesting feature of these results is that they are rather insensitive
to many of the parameters that describe the colliding system. 
In the scaling regime, 
the enhancement factor $r_2$ depends on the parameter 
$1/k$ only, which is assumed to be an intrinsic property of 
the excited partner. As a consequence, the enhancement must be 
roughly independent of the properties of the spectator partner, once
we study excitation of double GDR in the same nucleus.

This is just what has been observed in experiments: the values of $r_2^{exp}$
found for DGDR in $^{208}Pb$ projectile using different targets
$^{120}Sn, ^{165}Ho, ^{208}Pb,  ^{238}U$ \cite{B-208-Pb-1.33} 
are very close to each other and 
they correlate, within the error bars, with the value
\begin{equation}  \label{r2-208Pb}
r_2 (^{208}Pb) \simeq 1.33 \quad .
\end{equation}
According to the scaling prediction, this corresponds to the nonlinearity
parameter $k$ equal to 
\begin{equation}  \label{k-208Pb}
k (^{208}Pb) \simeq 1.5
\end{equation}
Numerical evaluation shows that the value of $\psi$ 
(see Eq.(\ref{PSI})) is close to $0.88$, thus, the value of the
required nonlinearity is a bit bigger, corresponding to 
$1/k \simeq 1$. 

The same picture was found in experiments on Coulomb desintegration 
of $^{197}Au$ target using various projectiles 
$^{20}Ne, ^{86}Kr, ^{197}Au,  ^{209}Bi$ \cite{AU-EXP-Au}.
Also, the similar conclusion of nearly constant value of $r_2$ 
in $^{208}Pb$ target with various projectiles
has been made in work \cite{B-208-Pb-LOW} (though for differing 
experimental set-up).

Below, we present the exact results for the cross sections
calculated according to Eqs.(\ref{sigman}), and (\ref{RESULT}) 
and with solving Eq.(\ref{RICCATI}) numerically.

The dependence of the enhancement factor $r_2=\sigma_2/\sigma^{harm}_2$ 
(\ref{PSI})
for the DGDR excitation on the strength of
the nonlinearity $k$ is shown on Fig. 1. for the process   
$^{208}Pb + ^{208}Pb$ (we use the case of bombarding energy 
$\varepsilon=0.64 GeV/$per nucleon). It is seen that the enhancement factor 
drops to unity at big values of $k$ (harmonic limit)
and grows at stronger nonlinearity. The scaling value of $r_2$
is also shown for comparison.

It is interesting to trace energy dependence of the enhancement factor. 

For typical projectile-target pairs and for 
moderate bombarding energies 
($\gamma \simeq 1.3$),
both $|\alpha|$ and $|\alpha^{harm}_{1}|$ are small as compared to 
unity and $|\alpha| \simeq |\alpha^{harm}_{1}|$.
The value of $R_2$ is well approximated by $\frac{2k+1}{2k}$, 
and this results in 
approximate scaling behavior
for the cross sections: one has  $r_2 = \sigma_2 / \sigma^{harm}_2
\simeq \frac{2 k +1}{2 k}$.
For reasonable values of $k$, the enhancement factor $r_2$ 
of the cross-section of Double Giant Resonance excitation 
can reach values 
$\sim 1.5 $ that is compatible with experimental data \cite{EMLING}.

Deviations from this simple scaling rule occur at both low and high
energies. At $\gamma \rightarrow 1$, adiabatic approximation 
is valid, and this yields to
$|\alpha| 
> |\alpha^{harm}_{1}|$.
Thus, $R_2 > R_2^{scaling} = (2 k+1)/(2k)$.
By contrast, at higher energies, the dynamical nonlinear effects 
tend to reduce the magnitude 
of $|\alpha|$, thus $|\alpha|/|\alpha^{harm}_{1}| <1$,
and $R_2 < R_2^{scaling}$. To sum up, the enhancement factor for the 
DGDR excitation cross section, $r_2 = \sigma_2 / \sigma^{harm}_2$ drops 
from $2-2.5$ (for low bombarding energies $\varepsilon \sim 100 MeV $ per 
nucleon) to
$1.2-1.3$ (for  $\varepsilon \sim 640-700 MeV$ per nucleon) 
while fixed value of
$k$ is used. 
On Fig.2., we plotted the value of the enhancement factor 
calculated numerically for the case
of $^{208}Pb+^{208}Pb$ process. The magnitude of nonlinearity is
kept fixed,  $k = 1.3$.

These results are is in correspondence with the experimentally 
observed trends and with 
microscopic models based on Axel-Brink concept \cite{hot-phonon1}.

To conclude, we presented here a simple model that accounts for the 
nonlinear effects in the transition probabilities for the excitation 
of multi-phonon Giant Dipole Resonances in Coulomb excitation 
via relativistic heavy ion collisions. The model is based on the 
group theoretical properties of the boson operators. It allows 
to construct the solution for the dynamics of the multi-phonon 
excitation within coupled-channel approach in terms of the 
generalized coherent states of the corresponding algebras.
The well known exactly solvable harmonic phonon model 
appears to be a limiting case of the present model when the 
nonlinearity parameter $1/k$ goes to zero. 
The main advantages of the limiting harmonic case (unrestricted
multiphonon basis, preservation of unitarity and possibility of
analytical treatment) remain present in our nonlinear
scheme.
Therefore, the model can be viewed as a natural extension of
the harmonic phonon model to include the nonlinear effects in 
a consistent way while keeping the model solvable.

At high enough projectile energies, the problem is simplified and 
the dynamics is governed by the generators of SU(1,1) group.
The problem of the excitation amplitudes is reduced to single
Riccati-type equation for the complex amplitude $\alpha$. 
Its solutions can be obtained by means of successive
approximations. In most cases relevant to applications, the perturbation
is weak enough to be considered in the first order; the explicit 
solution for the amplitude is then given by the 
modified Bessel functions. The probabilities to excite double-phonon
GDR appear to be enhanced by an approximately universal factor 
$(2 k + 1)/(2k)$; the same scaling is roughly valid for the  
cross sections. This can be viewed as a hint that the discrepancy 
between the measured cross-sections of double GDR and the harmonic
phonon calculations can be resolved within present nonlinear model
by means of using an appropriate value of the nonlinear parameter 
$1/k$ for a given nucleus. 
The experimental values of enhancement
of $\sigma_2$ with respect to the harmonic results for the excited 
$^{208}Pb$ nucleus are almost insensitive to the details of the
collision process.
On the other hand, the enhancement factor drops as the bombarding
energy grows. 
Though the present results are obtained in the 
``transverse approximation'' (neglecting the longitudinal response), 
they approximately hold in general case.
This is consistent with the data and 
gives results similar to those recently obtained in a possibly
different context, with a theory based on the concept of fluctuations
(damping) and the Brink-Axel mechanism 
\cite{hot-phonon1},\cite{hot-phonon2},\cite{CHP-HOT-3}.
It would be certainly worthwhile to establish possible connections
between the two approaches.

The work has been supported by FAPESP 
(Fundacao de Amparo a Pesquisa do Estado de Sao Paulo).

\newpage
\begin{center}
{\Large Figure Captions
}
\end{center}

\vspace{1cm}
Fig.1. Enhancement factor $r_2 = \sigma_2/\sigma_2^{harm}$
for the Double GDR excitation in 
$^{208}Pb + ^{208}Pb$ process at bombarding energy $\varepsilon=
640 MeV$/per nucleon as a function of the parameter $k$ (symbols, solid
curve is to guide the eye).
Dashed curve corresponds to the scaling $r_2 \simeq \frac{2 k +1}{2k}$.

Fig.2. Enhancement factor $r_2 = \sigma_2/\sigma_2^{harm}$
for the Double GDR excitation in 
$^{208}Pb + ^{208}Pb$ process as a function of relativistic factor 
$\gamma$ (symbols, solid curve is to guide the eye). 
The value of the nonlinear parameter is kept to be equal 
to $k = 1.3$. The scaling value (constant $\frac{2 k +1}{2k}$) 
is shown by dashed curve.

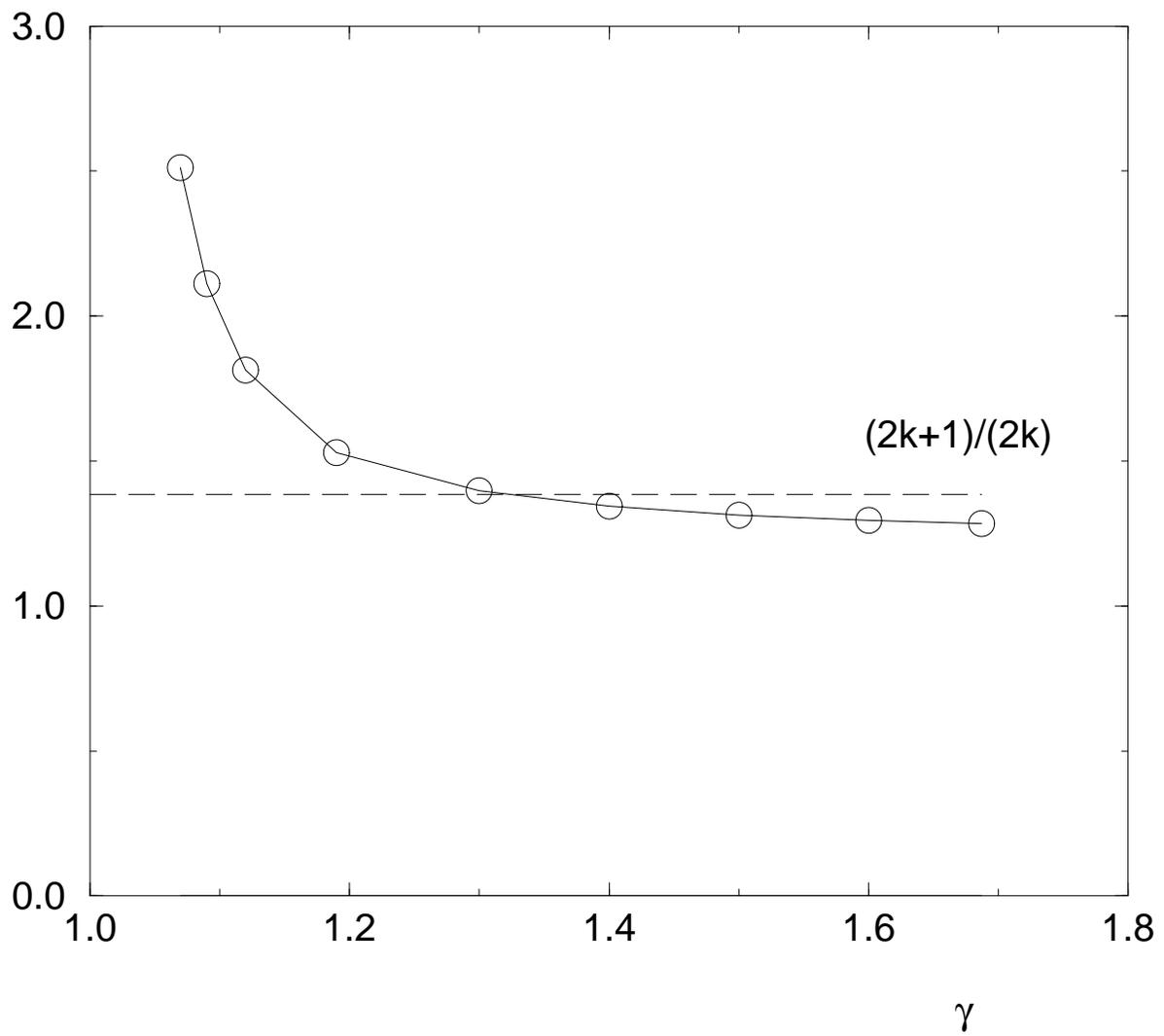
\end{document}